# Wide-field, high-resolution lensless on-chip microscopy via near-field blind ptychographic modulation†

Shaowei Jiang,‡a Jiakai Zhu,‡a Pengming Song,‡b Chengfei Guo,‡a Zichao Bian,a Ruihai Wang,a Yikun Huang,a Shiyao Wang,c He Zhang,a and Guoan Zheng *ab

We report a novel lensless on-chip microscopy platform based on near-field blind ptychographic modulation. In this platform, we place a thin diffuser in between the object and the image sensor for light wave modulation. By blindly scanning the unknown diffuser to different x-y positions, we acquire a sequence of modulated intensity images for quantitative object recovery. Different from previous ptychographic implementations, we employ a unit magnification configuration with a Fresnel number of ~50,000, which is orders of magnitude higher than previous ptychographic setups. The unit magnification configuration allows us to have the entire sensor area, 6.4 mm by 4.6 mm, as the imaging field of view. The ultra-high Fresnel number enables us to directly recover the positional shift of the diffuser in the phase retrieval process, addressing the positioning accuracy issue plagued in regular ptychographic experiments. In our implementation, we use a low-cost, DIY scanning stage to perform blind diffuser modulation. Precise mechanical scanning that is critical in conventional ptychography experiments is no longer needed in our setup. We further employ an up-sampling phase retrieval scheme to bypass the resolution limit set by the imager pixel size and demonstrate a half-pitch resolution of 0.78 μm. We validate the imaging performance via *in vitro* cell cultures, transparent and stained tissue sections, and a thick biological sample. We show that the recovered quantitative phase map can be used to perform effective cell segmentation of the dense yeast culture. We also demonstrate 3D digital refocusing of the thick biological sample based on the recovered wavefront. The reported platform provides a cost-effective and turnkey solution for large field-of-view, high-resolution, and quantitative on-chip microscopy. It is adaptable for a wide range of point-of-care-, global-health-, and telemedicine-related applications.

## Introduction

Optical microscope platform with high numerical aperture (NA) objective lens has long been used for microscale bioimaging. In contrast, lensless on-chip microscopy enables high-resolution imaging without using any lens. It can address the imaging needs of a wide range of lab-on-a-chip applications. Various lensless imaging platforms have been reported in the past years[1-20]. In lensless shadow imaging, the sample is placed on top of the active sensing area of the imager. Object intensity information can be recovered from the captured images via simple filtered back projection[6, 13-15]. Thanks to micron-level short distance between the object plane and the detector, this imaging modality has no stringent coherence requirement on the light source.

For other lensless imaging setups, the distance between the object plane and the detector is on the millimetre or centimetre scale. Coherent or partial coherent light source is typically needed for sample illumination. The captured intensity images represent coherent diffraction patterns of the complex object. With no phase-sensitive detector exists, the phase information of the light waves is lost in the acquisition process. This so-called phase problem is inherent to crystallographic, astronomical and optical imaging, or more generally, to all scattering experiments independent of the radiation used. For weakly scattering and sparse objects, the principle of digital in-line holography can be used to recover the complex object[1, 12]. Under more general consideration where a clean reference wave is not present, a phase retrieval process is needed for object recovery. Multiple object heights[4, 16], multiple incident angles[5, 8, 11], or multiple wavelengths[7, 17, 21] can be used in the acquisition process to introduce diversity to the phase retrieval process. Recently, we have also demonstrated the use of a translated unknown speckle pattern for lensless phase retrieval[9]. Different implementations of lensless imaging approaches can be found in the recent review papers[22, 23].

a. Department of Biomedical Engineering, University of Connecticut, Storrs, CT, 06269, USA.
b. Department of Electrical and Computer Engineering, University of Connecticut, Storrs, CT, 06269, USA.
c. Department of Department of Chemical & Biomolecular Engineering, University of Connecticut, Storrs, CT, 06269, USA
Email: guoan.zheng@uconn.edu

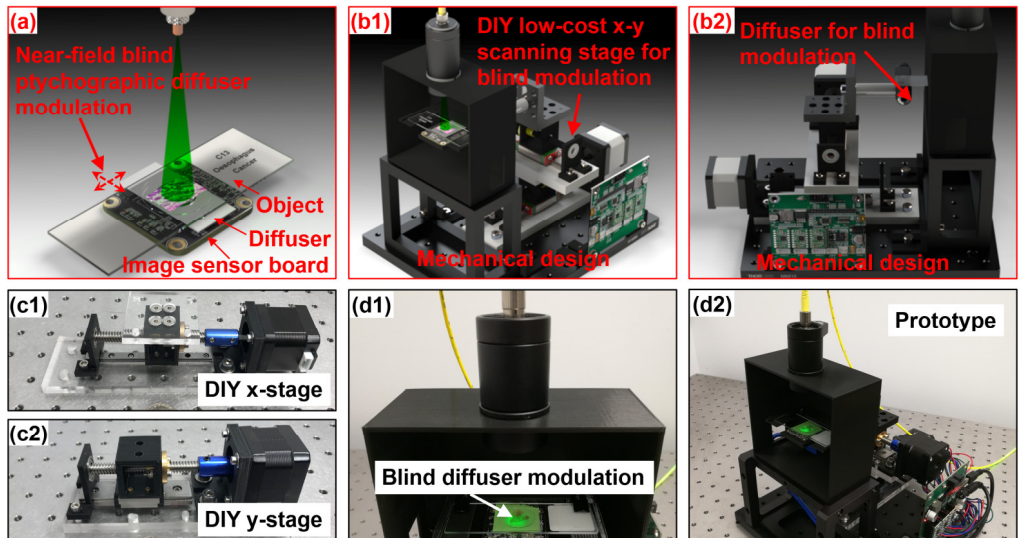

**Fig. 1** Lensless on-chip microscopy via near-field blind ptychographic diffuser modulation. (a) We place a thin diffuser in between the object and the image sensor for light wave modulation. The distance between the object and the image sensor is ~1 mm. (b) The design of the prototype platform. (c) The low-cost, DIY stages for blind diffuser modulation. (d) The cost-effective prototype setup. Movie S1 shows the 3D design of the reported platform. Movie S2 shows the operation of this prototype device.

Here we report a novel lensless on-chip microscopy approach for large field-of-view and high-resolution imaging. In this platform, we place a thin diffuser in between the object and the image sensor for light wave modulation. By blindly scanning the unknown diffuser to different x-y positions, we acquire a sequence of modulated intensity images for object recovery.

The reported platform shares its root with ptychography, a lensless imaging approach that was originally proposed for electron microscopy[24] and brought to fruition by Roderburg and colleagues[25]. Ptychography requires no phase modulation or interferometric measurements, enabling coherent diffraction imaging for visible light, X-rays, and electron[9-11, 24-32]. A typical ptychographic dataset is obtained by collecting a sequence of diffraction patterns in which the object is mechanically scanned through a spatially confined illumination probe beam. The confined probe beam limits the physical extent of the object for each diffraction pattern measurement and serves as a support constraint for the phase retrieval process. The acquired 2D images are then jointly processed into an estimate of the sample's complex transmission profile.

Different from previous lensless ptychographic implementations[9-11, 25, 27-31], we employ a unit magnification configuration with a Fresnel number of ~50,000. The unit magnification configuration allows us to have the entire sensor area, 6.4 mm by 4.6 mm, as the imaging field of view. The ultra-high Fresnel number enables us to directly recover the positional shift of the diffuser in the phase retrieval process, addressing the positioning accuracy issue plagued in regular ptychographic experiments[33, 34]. In our implementation, we use a low-cost, DIY scanning stage to perform blind diffuser modulation. Precise mechanical scanning that is critical in conventional ptychography experiments is no longer needed in our implementation. We further employ an up-sampling phase retrieval scheme to bypass the resolution limit set by the pixel size of the image sensor. Unlike the illumination-based approach[9, 28, 29], the recovered image of our platform only depends upon how the complex wavefront exits the sample. Therefore, the sample thickness becomes irrelevant during reconstruction. After recovery, we can propagate the complex wavefront to any position along the optical axis.

In the following sections, we will demonstrate the working principle of our lensless microscope platform and validate the performance of our prototype device. We show that the recovered quantitative phase map can be used to perform effective cell segmentation of the *in vitro* yeast culture. We also demonstrate 3D digital refocusing of a thick biological sample based on the recovered object wavefront. With a compact configuration and quantitative performance, we envision that the reported approach can find applications in a wide range of lab-on-a-chip platforms.

## Methods

**Near-field blind ptychographic modulation**

Figure 1(a) shows the working principle of our lensless imaging platform. In this platform, we use a 5-mW, 532-nm fiber-coupled laser diode for sample illumination. We make a thin diffuser by coating ~1 μm polystyrene beads on a coverslip and place it in between the object and the image sensor for light wave modulation. The resulting modulated intensity patterns are then captured by the image sensor. In this setup, the distance '*d*' between the object and the image sensor is ~1 mm, corresponding to a Fresnel number of ~50,000 (imaging area divided by '*d·wavelength*'). A large Fresnel number indicates little light diffraction from the object plane to the sensor plane. As such, the motion of the diffuser can be directly tracked from the captured raw images. To the best of our knowledge, the Fresnel number of our platform is orders of magnitude higher than previous ptychographic experiments.

In the acquisition process, we blindly scan the diffuser to different x-y positions and acquire the modulated intensity images using an image sensor (MT9J003 ON Semiconductor, 1.67 µm pixel size). We term this scheme 'near-field blind ptychographic modulation', where the word 'blind' has twofold implications. First, it implies the recovery of both the high-resolution complex object and the diffuser profile at the same time, similar to the recovery of both probe and object in blind ptychography[30, 35]. Secondly, it means we do not have any prior information of the positional shift of the diffuser. This second point, to the best of our knowledge, is new in ptychographic implementations, where precise positional information is typically required. The positional accuracy problem plagued in regular ptychography has generated considerable interest in developing algorithms to refine the positional shift during the phase retrieval process[33, 34]. However, these approaches may fail if a good initial guess of the shift is not available. Thanks to the ultrahigh Fresnel number, we can directly recover the positional shift with sub-pixel accuracy. Low-cost random scanning system can, thus, be used in the reported platform without any positioning accuracy requirement.

Figure 1(b) shows the design of our prototype device, where we use two low-cost, DIY mechanical stages to move the diffuser to x-y different positions. The motion step size is 1-2 µm in our implementation via motor micro-stepping. We note that the exact positional shift is treated as unknown in our experiment. Figure 1(c) shows our low-cost, DIY motion stages based on Nema-17 stepper motors, linear rail guide, and 8 mm lead screw (also refer to Figs. S1-S2 and Table S1 in Supplementary materials). Figure 1(d) shows the entire prototype platform, where we use the Arduino microcontroller to control the scanning the x-y stages. Movie S2 shows the operation of the prototype platform (the motion step size has been exaggerated for visualization).

**Imaging model and the phase retrieval process**

The forward imaging model of our platform can be expressed as

$$I_j(x,y) = \left| \begin{array}{c} O(x,y) * PSF_{free}(d_1) \cdot D(x-x_j, y-y_j) \\ * PSF_{free}(d_2) * PSF_{pixel} \end{array} \right|^2_{\downarrow M} \quad (1)$$

where $I_j(x,y)$ is the $j^{th}$ intensity measurement ($j = 1,2,...,J$), $O(x,y)$ is the complex exit wavefront of the object, $D(x,y)$ is the complex profile of the diffuser, $(x_j, y_j)$ is the $j^{th}$ positional shift of the diffuser, '·' stands for point-wise multiplication, and '*' denotes the convolution operation. In Eq. (1), $d_1$ is the distance between the object and the diffuser, and $d_2$ is the distance between the diffuser and the image sensor. We use $PSF_{free}(d)$ to model the point spread function (PSF) for free-space propagation over distance $d$, and $PSF_{pixel}$ to model the PSF of the pixel response. Due to the relatively large pixel size, the captured image is a down-sampled version of the diffraction pattern, and we use '$\downarrow M$' in the subscript of Eq. (1) to represent the down-sampling process ($M$ = 3 in our implementation). Based on all captured images $I_j$ with the diffuser scanned to different lateral positions $(x_j, y_j)$s, we aim to recover the complex exit wavefront of the object $O(x, y)$ and the diffuser profile $D(x, y)$.

---

**Algorithm outline**

**Input**: Raw images $I_j$ ($j = 1,2,\cdots,J$) with the translational shift of the diffuser
**Output**: High-resolution object $O(x,y)$ and the diffuser profile $D(x,y)$

1. Calculate the translational shift $(x_j, y_j)$ of the diffuser using cross-correlation
2. Initialize $O(x,y)$ and $D(x,y)$
3. $O_D(x,y) = PSF_{free}(d_1) * O(x,y)$ % Propagate the wavefront to the diffuser
4. **for** $n$= 1: $N$ (different iterations)
5.     **for** $j$= 1: $J$ (different captured images)
6.         $D_j(x,y) = D(x - x_j, y - y_j)$ % Shift the diffuser to $(x_j, y_j)$ position
7.         $\varphi_j(x,y) = O_D(x,y) \cdot D_j(x,y)$
8.         $\psi_j(x,y) = PSF_{free}(d_2) * \varphi_j(x,y)$
9.         Update $\psi_j(x,y)$ using Eq. (2) % Up-sampled magnitude projection
10.        $\varphi'_j(x,y) = PSF_{free}(-d_2) * \psi'_j(x,y)$
11.        Update the object using Eq. (3)
12.        Update the diffuser using Eq. (4)
13.        $D(x,y) = D_j^{update}(x + x_j, y + y_j)$ % Shift back the updated diffuser
14.     **end**
15. **end**
16. $O(x,y) = PSF_{free}(-d_1) * O_D^{update}(x,y)$ % Propagate back to the object plane

**Fig. 2** The recovery process of the reported lensless on-chip microscopy platform.

The recovery process is shown in Fig. 2. We first recover the positional shift of the diffuser using cross-correlation. We then initialize the amplitude of the object by averaging all measurements. The diffuser profile is initialized to an all-one matrix. In the iterative reconstruction process, we propagate the object to the diffuser plane and obtain $\varphi_j$ in line 7. At the detector plane, we use the following equation to update the exit wave $\psi_j(x,y)$:

$$\psi'_j(x,y) = \psi_j(x,y) \left( \frac{\sqrt{I_j(x,y)_{\uparrow M}}}{\sqrt{|\psi_j(x,y)|^2 * ones(M,M)_{\downarrow M \uparrow M}}} \right) \quad (2)$$

In Eq. (2), the image sizes of $\psi_j(x,y)$ and $I_j(x,y)$ are different. If $I_j$ has a size of 100 by 100 pixels, $\psi_j$ will have 300 by 300 pixels (with an up-sampling factor $M$ =3). The term '$I_j(x,y)_{\uparrow M}$' represents the nearest-neighbor up-sampling of the captured image $I_j$. In the denominator of Eq. (2), we first convolute the term $|\psi_j(x,y)|^2$ with an average filter (M by M all-one matrix). We then perform $M$-times down-sampling followed by $M$-times nearest-neighbor up-sampling. This updating process enforces the intensity summation of every $M$ by $M$ small pixels equals the corresponding large pixel in the captured image[36]. The updating process of Eq. (2) is also the same as that in single-pixel imaging via alternating projection[37, 38]. With the updated $\psi'_j(x,y)$, we then update the object and the diffuser profiles via Eqs. (3) and (4)[39, 40]:

$$O_D^{update}(x,y) = O_D(x,y) + \frac{conj(D_j(x,y)) \cdot \{\varphi'_j(x,y) - \varphi_j(x,y)\}}{(1-\alpha_{obj})|D_j(x,y)|^2 + \alpha_{obj}|D_j(x,y)|^2_{max}} \quad (3)$$

$$D_j^{update}(x,y) = D_j(x,y) + \frac{conj(O_D(x,y)) \cdot \{\varphi'_j(x,y) - \varphi_j(x,y)\}}{(1-\alpha_{pt})|O_D(x,y)|^2 + \alpha_{pt}|O_D(x,y)|^2_{max}} \quad (4)$$

The phase retrieval process converges within 2-3 iterations in all experiments reported in this work. This is much faster than regular ptychographic experiments where hundreds of iterations are often needed. The fast convergence speed may be due to the use of ultrahigh Fresnel number in our setup. The processing time for 400 raw images with 1024 by 1024 pixels each is ~50 seconds for 2 iterations using a Dell XPS 8930 desktop computer.

## Results

**Resolution characterization**

We first validate the prototype device using a USAF resolution target in Fig. 3. Figure 3(a) shows the captured raw image of the resolution target, where the speckle feature comes from the diffuser modulation process. Figure 3(b) shows the recovered diffuser profile. Figure 3(c) shows our high-resolution recovery of the object, where we can resolve 0.78-µm linewidth of group 9, element 3 on the resolution target. We note that the scanning trajectory can be arbitrary. The lateral scanning step size for the diffuser is 1-2 µm in the acquisition process. No prior positional knowledge is needed for the recovery process.

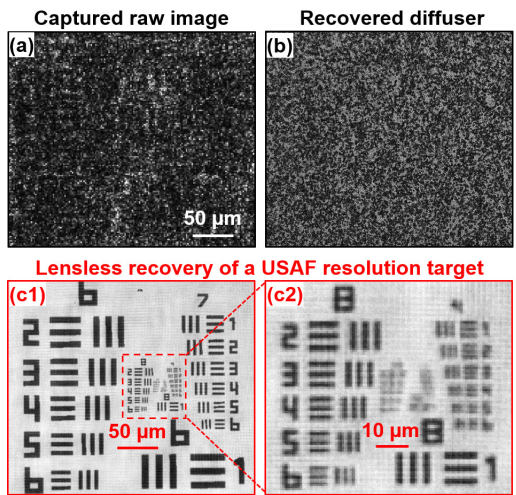

**Fig. 3** Validating the reported lensless approach using an amplitude resolution target. (a) The captured raw image. The recovered amplitude of the diffuser (b) and the object (c) based on 400 raw images.

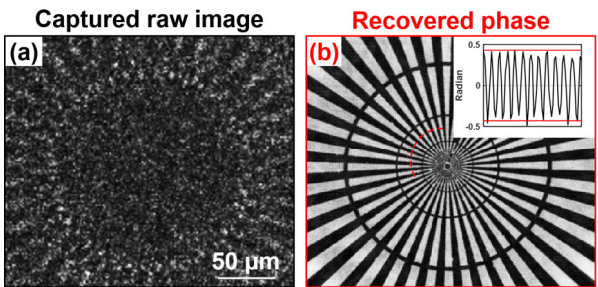

**Fig. 4** Validating the quantitative phase imaging nature of the reported approach. (a) The captured raw intensity image of the phase target. (b) The recovered phase image based on 400 raw images. The line trace of the red arc in (b).

**Quantitative phase imaging**

We also test our prototype using a quantitative phase target (Benchmark QPT) in Fig. 4. Figure 4(a) shows the captured raw image with diffuser modulation. Figure 4(b) shows the recovered quantitative phase using the reported platform. The line profile across the red dash arc in Fig. 4(b) is plotted in the inset. The recovered phase is in a good agreement with the ground-truth of the phase target (two red lines in the inset of Fig. 4(b)). We have also tested the phase imaging performance using an unstained mouse kidney slide (Fig. 5(a)) and transparent U87 cell culture (Fig. 5(b)). Figure 5(a1) and 5(b1) show the captured raw images. Figure 5(a2) and 5(b2) show the recovered quantitative phase of the two samples. Movie S3 shows the phase retrieval process of the unstained mouse kidney slide. Figures S3 and S4 show the recovered full field-of-view phase images of HeLa cell and U87 cell cultures. The quantitative phase imaging capability of the reported platform offers a label-free solution for cell-assay-related applications[41].

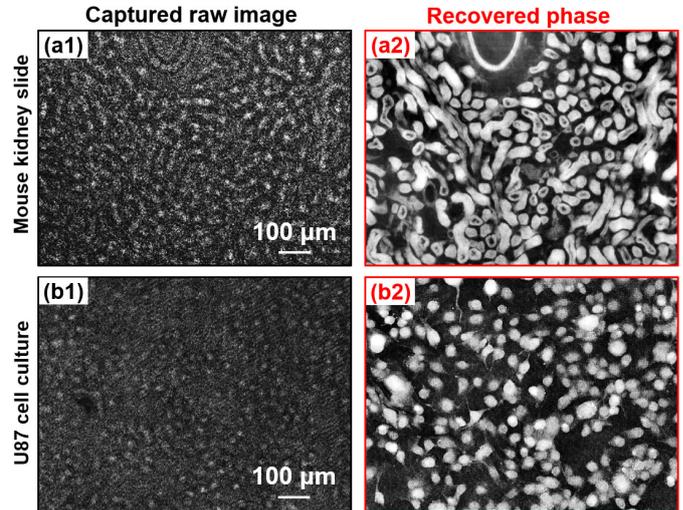

**Fig. 5** Testing of the reported approach using an unstained mouse kidney slide (a) and U87 cell culture (b). In this experiment, we directly use the recovered diffuser profile from the previous experiment. (a1), (b1) The captured raw images. (a2), (b2) The recovered phase profiles using 100 raw images. Movie S3 shows the recovery process.

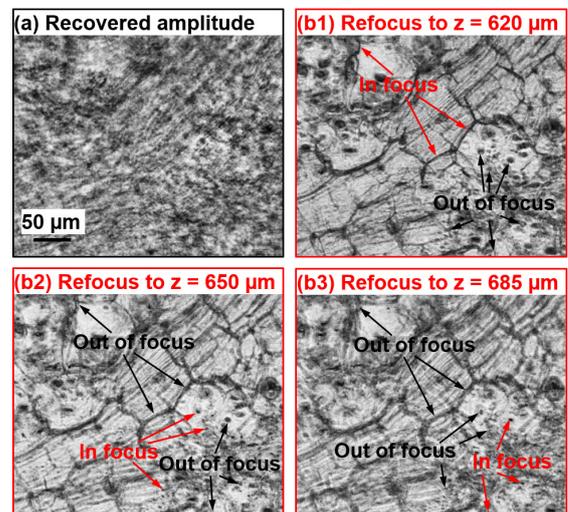

**Fig. 6** Testing the reported platform using a thick potato sample. (a) The recovered amplitude of the object's exit wavefront. (b) The three images after digitally propagating to three different axial positions. Movie S4 shows the digital propagation process.

Another advantage of the reported platform is to perform diffuser modulation at the detection path. Different from illumination-based approach[9, 28, 29], the recovered image of our platform only depends upon how the complex wavefront exits the sample[40, 42, 43]. Therefore, the sample thickness becomes irrelevant during reconstruction. After recovery, we can propagate the complex wavefront to any position along the optical axis. We validate this point using a thick potato sample in Fig. 6. Figure 6(a) shows the recovered amplitude of the object exit wavefront. Figure 6(b1)-6(b3) show the recovered object amplitude after digitally propagating to z = 620 μm, z = 650 μm, and z = 685 μm. The cell walls are in focus in Fig. 6(b1) and the organelles are in focus in Fig. 6(b2)-6(b3).

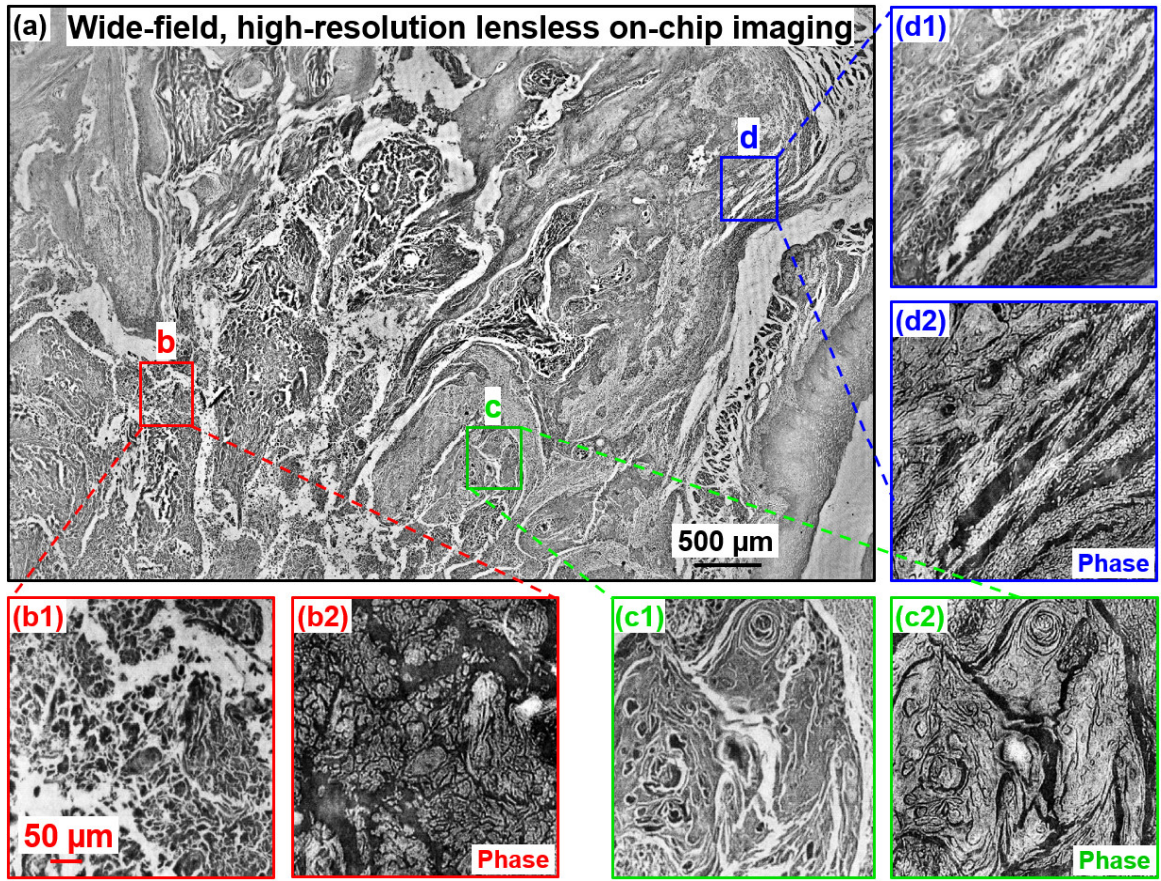

**Fig. 7** Wide field-of-view, high-resolution imaging of a stained esophagus cancer slide. (a) The full field-of-view image of the recovered object amplitude. (b)-(d) Magnified views of regions (b)-(d). Also refer to Fig. S5.

**High-resolution lensless imaging over a large field-of-view**

In many lab-on-a-chip applications, it is important to achieve both high-resolution and large field-of-view at the same time. One example is the detection of circulating tumor cells using lab-on-a-chip devices[44, 45], which often require microscopic imaging over a large field-of-view. The unit magnification configuration of the reported platform allows us to have the entire sensor area, 6.4 mm by 4.6 mm, as the imaging field of view. We validate the large field-of-view and high-resolution imaging performance using a stained esophagus cancer slide in Fig. 7. Figure 7(a) shows the full field of view of the recovered amplitude of the object. Figure 7(b)-7(d) show the magnified views of the recovered amplitude and phase of the three highlighted regions in Fig. 7(a). Figure S5 shows the comparison between the lensless images and the images captured by a standard light microscope.

**Effective automatic cell segmentation over a large field-of-view**

Conventional microscopic techniques like phase contrast or differential interference contrast suffer from shadow-cast artifacts which make automatic cell segmentation challenging. The quantitative phase map acquired by the reported lensless imaging platform enables an effective solution for rapid label-free cell segmentation.

In Fig. 8, we test the prototype platform for automatic cell segmentation using *in vitro* yeast culture. For yeast preparation, 2 ml of yeast extract peptone dextrose (YPD) medium was inoculated with a colony isolated yeast from a fresh YPD agar plate. The culture was incubated in culture tube overnight at 30°C with 150 rpm. On the following day, a fresh 1-ml YPD culture was prepared by inoculating 1 ml of YPD with 0.1 ml of overnight culture and incubated, with gentle agitation, for 1 hour at 30°C. To culture the yeast on the glass slides, we made a thin YPD solid agar layer on top of a cover glass. A 20 μl aliquot of yeast suspension was then added onto the YPD agar layer. The yeast culture on the cover glass was incubated at 30°C for 8 hours before we placed it in our lensless on-chip microscopy platform for image acquisition. Figure 8(a) shows the recovered phase image of the entire field of view. Figure 8(b) shows the magnified view of the phase image. Figure 8(c) and 8(e) show the corresponding raw image and the diffuser amplitude profile.

We use the following four steps to perform automatic cell segmentation in our study. First, we binarize the phase image by setting a threshold value that is inferred from histogram analysis. As such, we separate the yeast cells from the background. Second, we perform seed-point extraction based on local maxima detection. Third, we perform watershed transform to perform cell segmentation. Fourth, we refine the segmentation result based on the prior information of the cell size. In this refinement process, we first identify the segmented regions that are at least two times larger than the average cell area. We then enhance the image contrast of those regions and repeat steps 2 and 3. Figure 8(d) shows the result of automatic cell segmentation (different segmented cells are coded with different colours). For a 200-µm square region, there are 1550 cells via our automatic segmentation approach discussed above. For the same region, we have an average of 1568 cells via manual counting by three persons. The difference between the two is only about 1%, validating the effectiveness of the reported segmentation method using the quantitative phase map. For the entire field-of-view, the total number of yeast cells is 641,690 via our automatic cell segmentation approach and the processing time is 15 seconds. The reported automatic cell segmentation over a large field-of-view may find a wide range of applications from drug discovery to single-cell biology.

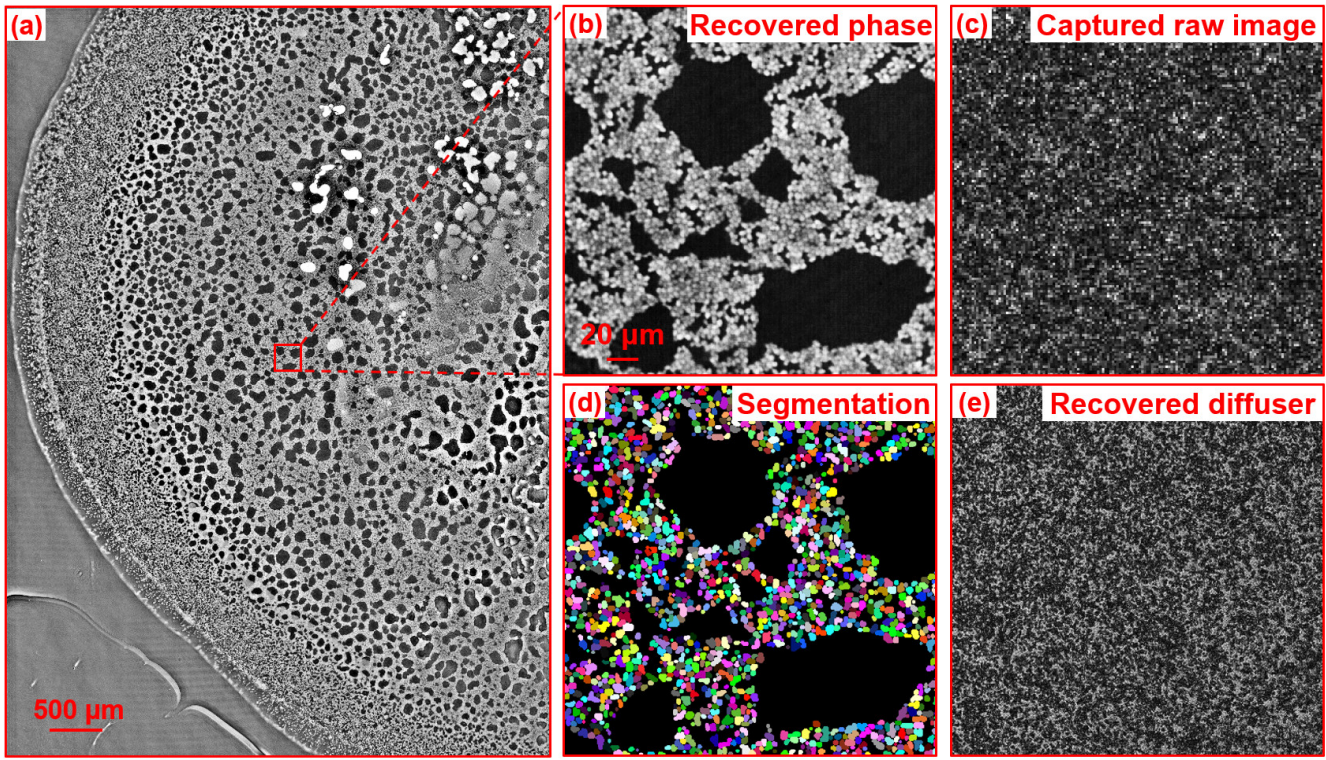

**Fig. 8** Wide field-of-view, high-resolution imaging of *in vitro* confluent yeast culture. (a) The recovered phase image of the yeast culture. (b) The magnified view of the yeast culture. The size of individual yeast cell is 3-4 µm. (c) The captured raw image corresponding to region shown in (b). (d) The segmentation result based on the proposed four-step procedure. Different cells are coded with different colours. (e) The recovered diffuser profile corresponding to region shown in (b). Also refer to Figs. S3-S4 for the recovered phase images of HeLa cell and U87 cell cultures.

## Discussions and conclusions

In summary, we report a novel lensless on-chip imaging platform for wide-field, high-resolution microscopy. In this platform, we place a thin diffuser in between the object and the image sensor for light wave modulation. By blindly scanning the unknown diffuser to different x-y positions, we acquire a sequence of modulated intensity images for object recovery. We demonstrate a half-pitch resolution of 0.78 µm and test the imaging performance using various confluent biological samples. We also demonstrate effective automatic cell segmentation based on the quantitative phase map generated by the reported platform.

There are several unique advantages of the reported platform. First, different from previous ptychographic implementations, the unit magnification configuration allows us to have the entire sensor area, 6.4 mm by 4.6 mm, as the imaging field of view. Second, thanks to the ultra-high Fresnel number, the speckle feature from the diffuser can be clearly resolved from the captured intensity images. Therefore, we can directly recover the unknown positional shifts via image cross-correlation. Precise mechanical scanning that is critical in conventional ptychography experiments is no longer needed in our implementation. Third, different from the pixel super-resolution technique employed in conventional lensless on-chip imaging setups[3, 13, 15, 16, 22], the reported platform can achieve sub-pixel resolution using an up-sampling phase retrieval scheme. Fourth, since we modulate the light wave at the detection path, the recovered image only depends upon how the complex wavefront exits the sample. After recovery, we can propagate the complex wavefront to any position along the optical axis. Lastly, our phase retrieval process typically converges within 2-3 iterations, which are much faster than regular ptychographic implementations. This may be due to the ultrahigh Fresnel number of our setup. Further research along this line is highly desired.

## Contributions

S. J., J. Z., P. S., C. G. contributed equally to the work and develop the reported prototype platform. G. Z. conceived the idea and planned the study. H. Z. performed the initial test. Z. B., S. W., Y. H. prepared the cell culture samples. All authors contributed to writing and correcting the manuscript.

## Conflicts of interest

There are no conflicts to declare.

## Acknowledgments

G. Z. acknowledges the support of the National Science Foundation 1510077, 1555986, and 1700941.

## Notes and references